\documentstyle[preprint,aps,psfig,eqsecnum]{revtex}
\begin{document}
\draft
\tightenlines
\preprint{\vbox{\hbox{U. of Iowa 97-2501; CERN 97-251}}}
\title{A Guide to Precision Calculations in Dyson's Hierarchical
Scalar Field Theory}
\author{J. J. Godina\\
{\it Dep. de Fis. , CINVESTAV-IPN , Ap. Post. 14-740, Mexico, D.F. 07000\\
and\\
Dpt. of Physics and Astr., Univ. of Iowa, Iowa City, Iowa 52246, USA}}
\author{Y. Meurice\cite{byline} and M. B. Oktay\cite{byline}\\ 
{\it CERN, 1211 Geneva 23, Switzerland\\ 
and\\ 
Dpt. of Physics and Astr., Univ. of Iowa, Iowa City, Iowa 52246, USA}}
\author{S. Niermann\\
{\it Dpt. of Physics and Astr., Univ. of Iowa, Iowa City, Iowa 52246, USA}}
\maketitle
\begin{abstract}
The goal of this article is to provide a practical method  to 
calculate, in a scalar theory, 
accurate numerical values of the renormalized quantities 
which could be used to test any kind of approximate calculation.
We use finite truncations of the Fourier transform of the
recursion formula
for Dyson's hierarchical model in the symmetric phase
to perform high-precision calculations of the 
unsubtracted Green's functions at zero momentum
in dimension 3, 4, and 5. We use the well-known correspondence
between statistical mechanics and field theory
in which the large cut-off limit 
is obtained by letting $\beta$ reach a critical value $\beta_c$ 
(with up to 16 significant digits in our actual calculations).
We show that the round-off errors on 
the magnetic susceptibility
grow like $(\beta_c -\beta)^{-1}$ near criticality.
We show that
the systematic
errors (finite truncations and volume) 
can be controlled with an exponential 
precision and reduced to a level
lower than the numerical errors.
We justify the use of the truncation for 
calculations of the high-temperature
expansion. 
We calculate
the dimensionless renormalized coupling constant corresponding to the 
4-point function and show that 
when $\beta \rightarrow \beta_c$, this quantity tends to a fixed value 
which can be determined accurately
when $D=3$ (hyperscaling holds),
and goes to zero like $(Ln(\beta_c -\beta))^{-1}$ 
when $D=4$. 
\end{abstract}
\newpage
\section{Introduction}
\label{sec:intro}

Finding closed form, exact analytical solutions to difficult problems is
considered as a great achievement in theoretical physics.
In recent years, the development of fast computers and of easy electronic 
communications has enlarged the class of solutions which can 
be considered as completely satisfactory. Lengthy expressions can be 
manipulated symbolically or numerically and communicated to others
(for a concrete example, see for instance the Appendices of Ref. \cite{jmp}). 
Sometimes, the solution of a problem requires a combination of iterations
and expansions which can be performed with any desirable precision in a short 
amount of time using a friendly environment such as {\it Mathematica}.
An example of such a solution is the calculation of the spectrum of the 
one-dimensional quantum anharmonic oscillator described in Ref. \cite{anh}. 
In this example, even though no closed expression for the eigenvalues
and eigenfunctions is available, the beginning of the spectrum
can be obtained numerically with great precision and almost instantly
using {\it Mathematica}.
We could say that this problem is numerically solvable.

Scalar field theory on a Euclidean lattice is a difficult problem 
with many important
applications, such as the interactions of  strongly interacting
light particles (pions, kaons,...), 
the generation of mass in the standard model of elementary 
particles, and the theory
of critical phenomena. The renormalization group method\cite{wilson}
helps us to understand the nature of the continuum limit\cite{wilson2}
for such a model. However, in the case of short range
(nearest neighbor) interactions, we are still far away from 
the numerical solvability mentioned above. 

In order to solve scalar field theory, 
one needs an approximation scheme such that: a) the
zeroth-order approximation preserves the main qualitative features 
of the model, b) the zeroth-order approximation is analytically
or numerically solvable, and c) the zeroth-order approximation 
can be improved systematically and in a practically
implementable way. The fact that Wilson's approximate recursion formula 
satisfies the requirement a) is justified in Ref.\cite{wilson}.
The approximate recursion formula is an integral equation with
one variable which can be handled by standard numerical methods.
The main sources of errors are the finite number of points of 
integration and the parametrization of the behavior of the tails
of the functions integrated. The errors can be reduced by
reducing the size of the tails and increasing the number of points.
We found this approach time consuming and inefficient. However,
using the Fourier transform of the recursion formula, we found 
a natural approximate method with a fast implementation
and a control of the systematic 
errors which is better than exponential. This method
is the main calculational tool used and discussed in the present article.

The approximate recursion formula is closely related to the recursion
formula appearing in Dyson's hierarchical model\cite{dyson}. 
More precisely\cite{fam},
at fixed dimension, there exists a one parameter family of recursion
formulas 
which interpolates continuously between the two. 
Following Ref. \cite{fam}, we call this parameter $\zeta $.
Seen from a practical point of view in our calculation, the
extension to another value of $\zeta$ in the recursion formula 
amounts to 
changing one number in one line of a two page program.
Physical quantities
such as the critical exponents vary slowly when $\zeta$ is 
varied, but there is nothing that singles out a particular
value of $\zeta$. In the following, we specialize the discussion
and the numerical study
to the case of Dyson's model ($\zeta=1/D$) because this model
has been studied\cite{sinai,high,prl,osc} in great detail in the past.
Dyson's model is a well-defined lattice model and it admits the same 
kind of expansions (weak and strong coupling, large-$N$ etc.) 
as any other scalar model on a cubic lattice.
The basic recursion formula and the approximation methods are explained
in section \ref{sec:rec}.  

For simplicity, all the calculations done below
use an initial Ising measure. The infinite cut-off limit
is obtained by fine-tuning the only adjustable parameter, namely, 
the inverse temperature $\beta$.  
All the results will be expressed in terms of $\beta_c -\beta$.
The general procedure\cite{wilson,wilson2} 
which relates small values of $\beta_c -\beta$ to
large values of the UV cut-off is well-known and will not be repeated here.
From a calculational point of view, the discussion would be essentially
the same if instead we had fixed $\beta=1$ and considered 
initial measures depending on a cut-off and several bare parameters.

The rest of the presentation is based on the following 
empirical fact: the systematic errors due to finite volume 
fall exponentially with the number of iterations ($n_{max}$
and the systematic errors due to
finite 
dimensional truncation fall faster 
than exponentially with
the dimension of the truncated space ($l_{max}$).
In calculations involving double precision, 
one can, without encountering major difficulties,
choose $n_{max}$ and $l_{max}$
in such a way that if we increase these parameters further, no  
change is observed in the results. One can then 
first determine the critical temperature
and the numerical errors. 
We need to discuss these first,
because the numerical errors are a
fundamental limitation
whenever the precision of the arithmetic operations is fixed, and there 
is no point in trying to reduce the systematic errors much below the
numerical errors.

The phase structure of the approximated models is discussed in section 
\ref{sec:phase}.
We show that, for understandable reasons, there is nothing that can be 
identified as a low-temperature phase. We also give a practical method 
to identify the inverse of the critical temperature, denoted $\beta _c$,
with an 
optimal accuracy. In the following sections, it will always be assumed that 
we work in the symmetric phase ($\beta <\beta _c$).

The numerical errors on the magnetic susceptibility 
are studied in section \ref{sec:num}, where 
we show that the relative errors obey the approximate law
\begin{equation}
|{{\delta \chi }\over{\chi}}|\sim {{ \delta}\over{\beta _c - \beta}} \ ,
\end{equation}
where $\delta $ is 
the precision used to perform 
the arithmetical operations.
We also show that this law follows from an approximate renormalization
group calculation. As a by-product we obtain a numerical estimate of 
the critical exponent $\gamma$ in good agreement with the 
existing estimates\cite{prl,osc,epsi} for $D=3$. 
As a general remark, we have a much better quantitative control
on the details of renormalization group arguments in $D=3$ than 
in $D=4$, where confluent logarithmic singularities make the analysis 
more delicate.

We then discuss the systematic errors.
The volume dependence of the magnetic susceptibility is discussed
in section \ref{sec:vol}. We show that in order to calculate the 
susceptibility at an inverse temperature 
$\beta $, in $D$ dimensions, and with relative errors $\Delta $, one needs 
a number of lattice sites which is of the order of 
$(\Delta (\beta _c -\beta )^{\gamma})^{-D/2}$.
We then show that when the Fourier transform of the recursion
formula of Dyson's hierarchical model is projected into a finite dimensional
space of dimension $l_{max}$, the relative errors 
on the susceptibility associated with this truncation
decrease faster than $e^{-a \ l_{max}}$ for some positive number 
$a$. We show in section \ref{sec:ht} that similar results apply
to the high-temperature expansion of the susceptibility,
justifying the procedure used in Ref.\cite{osc}.

At this point, we have only discussed the 
susceptibility, or in other words, the 2-point function. 
The 4-point function is also provided by the calculational
method at no extra cost. However, the calculation of the 
corresponding renormalized coupling constant requires
a subtraction. In the discussion of the
errors on the susceptibility, we have explained that
while we are iterating the recursion formula we lose 
significant digits ``from
the right''. With the subtraction, we also lose significant digits
``from the left''. This is explained in section \ref{sec:lam},
where we calculate a ``dimensionless coupling constant'' 
inspired by the field theory definition of Ref. \cite{parisi}
and denoted $\lambda_4$ .

In the case $D=3$, we show that when $\beta \rightarrow \beta_c$, 
$\lambda_4$ tends to a fixed non-zero value that we were able to calculate
with 6 significant digits. This is very convincing evidence that 
hyperscaling holds for the model considered here. 
For the nearest-neighbor Ising
model on a 3-dimensional cubic lattice, it is very hard
to decide if hyperscaling
holds on the basis of Monte-Carlo simulations\cite{baker2}
or high-temperature expansion\cite{dis}. 
In the present case, it
is a short and straightforward calculation. 
This shows that it would be worth 
trying to interpolate perturbatively between Dyson's model and nearest 
neighbor models, transforming the qualitative approximation\cite{wilson}
into a quantitative approximation.
The accuracy of $\lambda_4$ decreases
when $D$ increases. However, we were able to obtain good
evidence that in $D=4$, $\lambda_4$ decrease like $(Ln(\beta_c -\beta))^{-1}$,
in good agreement with the behavior obtained with the 
field theory method\cite{brezin} at lowest order in perturbation theory.

Our results show that it is possible to calculate accurately, and
without major effort, the 
renormalized quantities 
which can be extracted from the 2- and 4-point functions.
These calculations can be done 
for an extended range of small values of $\beta_c -\beta $. 
In other words, for existing computers, requirement b) is
fulfilled by the hierarchical approximation provided that one
does not require too-small values of $\beta _c -\beta $.
Numerical solvability can sometimes completely change
our point of view regarding a problem. Taking the example of differential
equations, at a time when their numerical solutions were not 
achievable, one could dream of a perfectly deterministic
approach to natural phenomena in which everything could be known
once the rules of evolution and the initial conditions were given. 

This paper provides a calculational method
to anyone who would like to check an approximation method 
with accurate numerical results. 
As explained above, most of the approximation schemes which
apply to a scalar field theory with nearest neighbor interactions on a cubic
lattice also apply to Dyson's model.
We are presently
testing the validity
of several well-known but not easy-to-control
expansions: the renormalized perturbative expansion,
the loop expansion, and the large-$N$ expansion.

\section{The Recursion Formula and its Finite Dimensional Truncations}
\label{sec:rec}

In this section, we briefly describe Dyson's hierarchical model
and the methods used to calculate the average of arbitrary powers
of the zero momentum component of the scalar field.
The models considered here have $2^{n_{max}}$ sites.
We label the sites with $n_{max}$ 
indices $x_{n_{max}} ..... x_1$, each index
being 0 or 1.
In order to visualize the meaning of 
this notation, one can divide
the $2^{n_{max}}$ sites into two blocks, each containing $2^{n_{max}-1}$ 
sites. If $x_{n_{max}} =0$, the site is in
the first block, if $x_{n_{max}}=1$, the site is in the second block. 
Repeating this procedure $n_{max}$ times
(for the two blocks, their respective two sub-blocks, etc.),
we obtain an unambiguous labeling for each of the sites.
Two sites differing only by $x_1$ are in the same block of size 2.
We write the action as
\begin{equation}S=-{1 \over 2} \sum\limits_{n=1}^{n_{max}}({c \over 4})^n
\sum\limits_{x_{n_{max}},...,x_{n+1}}\ (\ \sum\limits_{x_n,....,x_1} \phi
_{(x_{n_{max}},....,x_1)} )^2\  \ .
\end{equation}
The index $n$, referred to as the ``level of interaction''
hereafter, corresponds to interactions of the total field in blocks
of size $2^n$.
The free
parameter $c$, controlling the strength
of the interactions, is set equal to $2^{1-2/D}$ in order to approximate
a nearest neighbor model in $D$-dimensions. In this article,
we will consider the cases $D=$ 3, 4, and 5.

The field $\phi_{(x_{n_{max}},....,x_1)}$ 
is integrated with a local measure $W_0 (\phi)$ which needs to
be specified. In the following, we use an 
Ising measure, $W_0(\phi) = \delta (\phi ^2 -1)$, which
takes only the values $\pm 1$. To the best of our knowledge, the results
presented do not depend on this specific choice and would also
apply, for instance to the case of Landau-Ginzburg measures
of the form $W_0(\phi) =  e^{-A\phi^2-B \phi ^4}$.

The integrations can be performed iteratively using the 
recursion formula 
\begin{equation}
W_{n+1}(\phi)=
{C_{n+1}\over 2}   e^{{\beta \over 2} ({c\over 4})^{n+1} \phi ^2}
\int d\phi ' W_n({{(\phi -\phi ')}\over 2})
W_n({{(\phi +\phi ')}\over 2}) \ ,
\end{equation}
where $C_{n+1}$ is a normalization factor which can be fixed
at our convenience.
The relation between this recursion formula and 
Wilson's  approximate recursion formula\cite{wilson}
is discussed in Ref.
\cite{fam}.
Introducing the Fourier representation
\begin{equation}
W_n(\phi)=\int {dk\over {2\pi}}e^{ik\phi} {\widehat W}_n(k) \ , 
\end{equation}
and a rescaling of the source $k$ by a factor $1/s$ at
each iteration, through the redefinition
\begin{equation}
R_n(k)={\widehat W}_n({k\over {s^n}}) \ ,
\end{equation}
the recursion formula becomes\cite{high}
\begin{equation}
R_{n+1}(k)=C_{n+1}\exp(-{1\over 2}\beta 
({c\over 4}s^2)^{n+1 }{{\partial ^2} \over 
{\partial k ^2}})(R_{n}({k\over s}))^2 \ . 
\label{eq:rec}\end{equation}
The rescaling operation
commutes with iterative integrations and the 
rescaling factor $s$
can be fixed at our convenience.

In the following, we fix the normalization constant $C_n$ in such way
that $R_n(0)=1$. $R_n(k)$ has then a direct probabilistic
interpretation.
If we call $M_n$ the total field $\sum \phi _x$ inside
blocks of side $2^n$, and $<...>_n $
the average calculated without taking into account the interactions
of level strictly larger than $n$ (or in other words, as if $n$
were equal to $n_{max}$), we can write
\begin{equation}
R_n(k)=\sum_{q=0}^{\infty}{{(-ik)^{2q}}\over{2q!}} 
{<(M_n)^{2q}>_n\over{s^{2qn}}} \ .
\label{eq:moments} \end{equation}
We see that the Fourier transform of the local measure 
obtained after $n$ iterations generates the 
zero-momentum Green's functions calculated with 
$2^n$ sites, and can thus be used to calculate 
the renormalized coupling constants at zero 
momentum. 

The choice of $s$ is a matter of convenience.
For calculations in the symmetric (high-temperature)
phase not too close to the critical temperature,
or for high-temperature expansions\cite{high,prl},
the choice $s=\sqrt{2}$ works well.
For calculations very close to the critical temperature, the 
choice $s={2 {c^{-{1\over 2}}}}$ prevents the appearance
of very large numbers.
In the following calculations,
we {\it always} use $s={2 {c^{-{1\over 2}}}}$.

In the following, the finite volume magnetic susceptibility is defined as 
\begin{equation}
\chi_n (\beta) = {<(M_n)^{2}>_n \over {2^{n}}} \ .
\end{equation}
With the rescaling $s={2 {c^{-{1\over 2}}}}$, we have the relation
\begin{equation}
\chi_n = -2 a_{n,1} ({2\over c})^n \ .
\end{equation}
The initial condition for the 
Ising measure is
$R_0=cos(k)$. For the Landau-Ginsburg measure, the coefficients in the 
$k-$expansion need to be  evaluated numerically.

In previous publications\cite{prl,osc}, we used 
the $\beta $ expansion of Eq. (\ref{eq:rec}) to calculate the 
high-temperature expansion of the magnetic susceptibility
of Dyson's hierarchical model up to order 800 with an Ising or a 
Landau-Ginzburg measure. 
The calculation of the large-order coefficients requires a lot
of computing time. We found that using a truncation in the expansion
in $k^2$ at order 50  could cut the computer time 
by a factor of order 100 while having  
effects on the values of the
coefficients which were smaller than the errors due to
numerical round-off. 

We consider the finite dimensional
approximations of degree $l_{max}$:
\begin{equation}
R_n(k)=1+a_{n,1}k^2 +a_{n,2}k^4+..... +a_{n,l_{max}}k^{2l_{max}} \ .
\label{eq:rn} \end{equation}
After each iteration, non-zero coefficients of higher 
order ($a_{n+1,l_{max}+1}$ etc.) are obtained, but 
not taken into account (i.e.  set to zero
as part of the approximation) in the next iteration.  
More explicitly, the recursion formula for the $a_{n,m}$ reads:
\begin{equation}
a_{n+1,m}={{\sum_{l=m}^{l_{max}}(\sum_{p+q=l} a_{n,p}a_{n,q})
{{(2l)!}\over {(l-m)!(2m)!}}
({c\over 4})^l
(-{1\over 2}\beta)^{l-m}}\over
{\sum_{l=0}^{l_{max}}(\sum_{p+q=l} a_{n,p}a_{n,q}){{(2l)!}\over {l!}}
({c\over 4})^l
(-{1\over 2}\beta)^{l}}} \ .
\end{equation}

\section{The Phase Structure of the Approximated Models}
\label{sec:phase}

From a field theoretic point of view, an important
feature of the hierarchical model is its second order 
phase transition. Our first task will be to identify $\beta _c$
for the well-studied\cite{prl,num} case of an Ising measure.
The truncation described above can be used to calculate\cite{high} 
{\it exactly} the high-temperature
expansion of the magnetic susceptibility up to order
$l_{max} -1$, and one can think about the truncation as a partially
re-summed high-temperature expansion.
It seems thus unlikely that  
when $l_{max}$ becomes large, we would obtain sensible 
results when $\beta > \beta _c$. This guess is confirmed by
empirical numerical experiments at fixed $\beta $. 

The truncated recursion formula
shows very clearly the existence of a high-temperature
phase where for $\beta$ smaller than $\beta _c$ and $n$ large enough, we have
the scaling law
\begin{equation}
a_{n,1} \propto 2^{-{2n\over D}} \ .
\end{equation} 
Given the choice 
of scaling factor $s={2 {c^{-{1\over 2}}}}$ 
and the definition of $c$
discussed in section \ref{sec:rec}, this scaling is
equivalent to the ``central limit'' behavior
\begin{equation}
<(M_n)^{2q}>_n \propto 2^{nq} \ .
\label{eq:central} \end{equation}
This situation is characterized by  ratios $a_{n+1,1}/a_{n,1}$
reaching the value $c/2=2^{-{2\over D}}$.

\begin{figure}
\vskip30pt
\centerline{\psfig{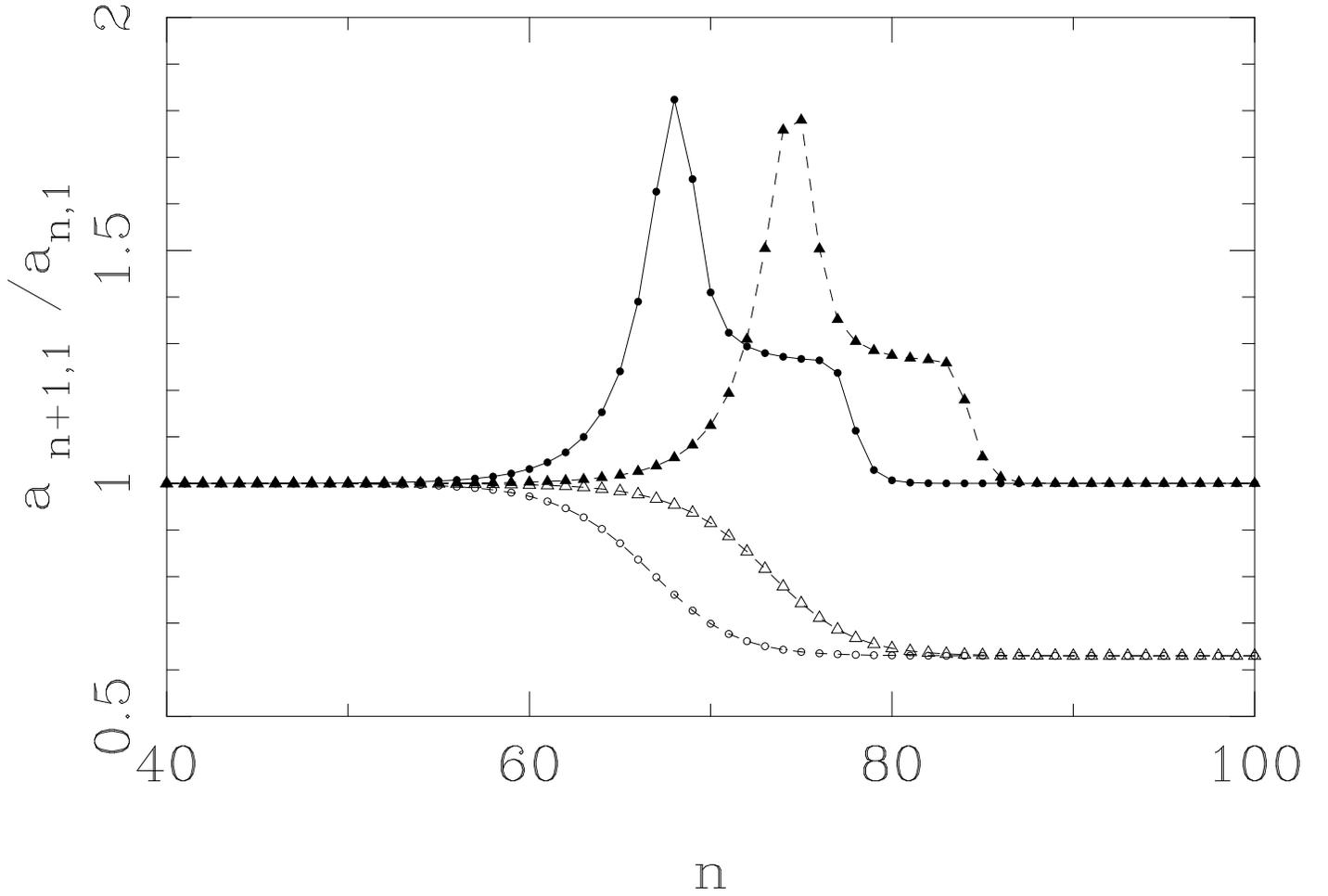}}
\vskip50pt
\caption{ 
$a_{n+1,1}/a_{n,1}$ versus $n$ for $\beta=\beta_c-10^{-10}$ 
(empty circles), $\beta=\beta_c-10^{-11}$ (empty triangles), 
$\beta=\beta_c+10^{-10}$ (filled circles) and $\beta=\beta_c+10^{-11}$
(filled triangles). }
\label{phase}
\end{figure}

On the other hand, when $\beta$ is increased
sufficiently, 
there is a sudden change of behavior. However,
there is nothing that we can identify with a low-temperature
phase\cite{parisi}. In all the cases that we have considered, the
change is signaled by the fact that when $n$ increases,
the ratio $a_{n+1,1}/a_{n,1}$ becomes larger than 1, and then starts making
unpredictable changes, {\it never} returning to any kind
of behavior like Eq. (\ref{eq:central}) or the behavior characteristic
of the low temperature phase, namely
\begin{equation}
<(M_n)^{2q}>_n \propto 2^{n2q} \ .
\end{equation}
The ``irreversibility'' of this process allows us to
identify unambiguously $\beta _c$, in the sense that a 
calculation at finite $n$ gives upper and lower bounds on $\beta_c$.
By increasing $n$, we can obtain sharper bounds.
This procedure is illustrated for $D=3$ 
in Fig. \ref{phase}.

We see that a calculation
for $n$ up to 50 allows us to resolve the 10-th digit of $\beta_c$, and
a calculation for $n$ up to 60 allows us to resolve the 11-th digit.
Proceeding similarly,
we can determine the  numerical value of $\beta _c$ 
with as many significant digits as the computational method allows.
Double precision Fortran  calculations  made with $l_{max}$ = 80  
are reported below
for the Ising model in 3, 4, and 5 dimensions. 
The results are in agreement with the bounds found in ref. \cite{num}
with independent (and exact) calculational methods.
The third  column gives the
minimal value of $n$ which allows a resolution of the 16 
significant digits.
\narrowtext
\begin{quasitable}
\begin{tabular}{cccc}
$D$& $\beta _c$&$n_{min}$&$l_{{max}_{min}}$\\
\tableline
3&1.179030170446270&102&32\\
4&0.6654955715318593&111&43\\
5&0.4569633006170210&132&45
\end{tabular}
\end{quasitable}
\begin{equation}
\label{eq:crit}
\end{equation}

Subsequently, $l_{max}$ was lowered by small steps until the value of $\beta _c$
changed. This experiment shows that the change occurs at values much smaller
than 80. In  the fourth column of Table 1, we give the minimal value of 
$l_{max} $ such that the stable value of $\beta_c$ with 16 
significant digits can be reached.

One may wonder if the precise value of $\beta _c$ is dependent on the 
numerical aspects of the calculation such as the round-off errors, 
which will be discussed
in the next section. To settle this question, we have used methods which
perform the arithmetic operations differently (these methods are explained 
in detail in the next section)
and found the very same values of $\beta _c$.
In conclusion, we have found a reasonably robust value of $\beta _c$
which is consistent with existing results. 

We have calculated $\beta _c^{-1} $ for dimensions 
much larger than 4. The results are displayed in Fig.~\ref{invbetc}.
We see that this quantity grows linearly with the dimension.
The explicit calculation of the high-temperature expansion\cite{jmp}
suggests that when $D\rightarrow \infty$, i.e. when $c\rightarrow 2$, we have
\begin{equation}
b_m\simeq ({2c\over{(4-c)(2-c)}})^m \ ,
\end{equation}
which implies that
\begin{equation}
\beta _c ^{-1} \simeq {D\over{2Ln2}} \ .
\end{equation}
This estimate of the slope is in good agreement with the data. We found a 
slope of 0.714, while $(2Ln(2))^{-1}=0.721$.
Next we will study
the various sources of error occurring when one approaches $\beta_c $ from
below.

\section{The Round-off Errors}
\label{sec:num}

Round-off errors can play an important role when recursive methods 
are used, because they may grow faster than the improvement of the 
results due to the repeated use of the method.
For this reason, we have studied them with three
independent methods. By independent, we mean that the arithmetic
is performed in a completely uncorrelated fashion.

We have compared our original Fortran calculation on a DEC-alpha with three
other calculations. The first one was the same program run on a MIPS.
The second one was a Mathematica program where a higher precision
in the arithmetic operations was set. 
The precision was adjusted in such a way that the susceptibility
was obtained correctly with 16 significant digits.
Thirdly, we have compared the 
calculation with the one obtained with a slightly different rescaling,
namely $s={1.98 {c^{-{1\over 2}}}}$, 
a method already used in Ref. \cite{prl,osc}.
All these calculations were performed with $l_{max}=60$, which is beyond
what we need (see next section).

The relative 
differences in the finite volume susceptibility are shown 
in Fig. \ref{3err},
for $D=3$ and $\beta _c -\beta = 10^{-11}$. The figure shows clearly that
the three types of discrepancies are essentially the same. Since the 
three types of errors are uncorrelated, we can identify them with 
the round-off errors and calculate them with the most convenient method.
Using the third method, we have calculated the round-off errors for 
various values of $\beta $. 
In all the cases, the logarithm of the relative error grows linearly
at the beginning and then stabilizes at a constant value.
The period of linear growth corresponds roughly to the iterations
where $\chi _{n+1} \simeq {2^{2\over D}} \chi_n$. For larger $n$, 
the value of the susceptibility stabilizes,
with changes decreasing by a factor $2^{-{2\over D}}$ at each step. 
During this second stage, the numerical errors do not grow significantly.

We now proceed to discuss the asymptotic values of the errors, in other 
words, the stable value they reach for $n$ sufficiently large.
We have collected these values for various temperatures and
$D=$3, 4, and 5 in Fig. \ref{dsus345}. This shows that the relative error
is in good approximation $10^{-16} (\beta_c - \beta )^{-1}$,
independently of $D$.

These empirical results have a simple explanation in terms of 
the linearized theory. Suppose that $\delta $ is a
typical round-off error in a calculation (e.g. $10^{-16}$), and that
$\lambda $ is the largest eigenvalue of the linearized renormalization
group transformation near a given fixed point. One expects the 
numerical error on $a_{n,1}$ to be of the order $\lambda ^n \delta$.
With the rescaling used in this paper, this means that the errors
on the susceptibility are of the order
\begin{equation}
|\delta \chi _n|\sim \lambda ^n \delta ({2\over c})^n \ .
\end{equation}
Now for $n$ such that 
$\lambda ^n \sim (\beta _c - \beta )^{-1}$, we have $\chi \sim 
({2\over c})^n$. Plugging this into the previous equation, we get
\begin{equation}
|{{\delta \chi }\over{\chi}}|\sim {{ \delta}\over{\beta _c - \beta}} \ ,
\end{equation}
which is the empirical result found above.

For $D=3$, one can check the details of the above argument and, as a
by-product, obtain an estimate of $\lambda $.
The slope of the increasing part of Fig.~\ref{3err} 
is approximately 0.154, with
an estimated error of order 0.001. This implies a value $10^{0.154}=1.427$
for $\lambda$. Using the usual formula for the critical exponent,
\begin{equation}
\gamma = {Ln({2\over c})\over Ln(\lambda)} \ ,
\end{equation}
we obtain the value $\gamma = 1.30$, in good agreement with existing 
estimates\cite{prl,osc,epsi}.

For $D=4$, the same procedure gives an exponent which is too high by
about 3 percent (compared to the trivial value). This is a typical
error when one does not take into account the marginal direction
and the (related) 
confluent logarithmic singularity in the susceptibility.
  
\section{Volume Effects}
\label{sec:vol}

In the two previous sections, we developed a qualitative understanding
regarding the finite volume susceptibility $\chi_n$, or in other 
words regarding the 
way the susceptibility depends on the number of iterations.
Volume effects can be important in the determination of 
the critical exponents. 
For instance, in Ref. \cite{num}, {\it exact} calculation with almost
a million sites gave errors of more than 10 percent in the exponent $\gamma $.
We are now ready to get a better quantitative understanding of these
effects.

If we consider the evolution of $\chi_n (\beta)$ when $n$ increases, with
$\beta $ fixed slightly below $\beta _c$, we see from Fig. \ref{phase} 
that when we are close to
the fixed point, $\chi _{n+1} \simeq {2^{2\over D}} \chi_n$. This lasts
until the right order of magnitude ( $\sim (\beta _c - \beta )^ {-\gamma}$ )
is reached. For larger $n$, the value of the susceptibility stabilizes,
with errors decreasing at each step. 
In this second regime, the measure becomes asymptotically Gaussian,
and one can estimate the change in $\chi_n$ from the change in the $k^2$ term.
From the basic formula (2.5), one gets the estimate for the relative change:
\begin{equation}
\Delta _n =\vert{ {\chi_{n+1} -\chi _n}\over {\chi _n}}\vert \propto 
2^{-{2\over D}n} \chi_n \ .
\label{eq:delta} \end{equation}
From these considerations, we find the number $n(\beta , \Delta)$
of iterations necessary to calculate the susceptibility at fixed $\beta$, with
a relative precision $\Delta $ (defined as in Eq. (\ref{eq:delta})) :
\begin{equation}
n(\beta , \Delta )=-({DLn(10)\over{2Ln(2)}})(Log _{10}(\Delta)+\gamma Log_{10}
(\beta _c -\beta)) \ .
\label{eq:nbeta} \end{equation}
The comparison with a numerical calculation  where we
required  $\Delta = 10^{-15}$ 
is given in Fig. \ref{vol}
for $D=$3, 4, and 5. The agreement with the estimate of Eq. (\ref{eq:nbeta})
with $\gamma$ = 1.3 (1.0) for $D=3$ (4 and 5), is quite good.

The fact that we were able to stabilize sixteen digits of the susceptibility 
does not mean that the results have sixteen digit accuracy. The
asymptotic stability of the numerical results comes from the fact that the 
r.\ h.\ s.\ of Eq. (\ref{eq:delta}) will go to zero whenever $\chi_ n $ quits 
growing. This occurs independently of the fact that numerical 
errors may occur while $\chi _ n$ builds up its bulk value.
Consequently, a more realistic approach would be to require a precision
consistent with the round-off errors discussed in the previous 
section. Imposing a temperature-dependent 
requirement   $\Delta = 10^{-16} (\beta_c - \beta )^{-1}$, we obtain 
values of $n$ shown in Fig. \ref{vol}. 

\section{Controlling the Effects of Finite Dimensional Truncations}
\label{sec:ldep}

In this section, we study the $l_{max}$ dependence
of the magnetic susceptibility
for $\beta < \beta _c$.
For notational purposes, we call $\chi^{(l)}$ the susceptibility corresponding
to a given value $l_{max}=l$. For each calculation, the value of $n_{max}$
has been increased until no change could be observed. The results are 
displayed in Fig. \ref{ldep} for $\beta_c -\beta\ = \ 10^{-8}$. For
low $l$, $\chi^{(l)}$ grows at a not-very regular rate
and within the bounds $1<\chi ^{(l+1)}/ \chi ^{(l)}<10 $. 
When $l$ gets close to 20, $\chi^{(l)}$ starts stabilizing with a precision
which seems to be exponential. For instance, for $D=3$, the relative errors
fall approximately like $10^{-0.6 l}$. This exponential rate is based on the 
assumption that the logarithm of the relative errors falls linearly.
However, a  closer look shows that it falls slightly faster. 
This is illustrated in Fig. \ref{ldepcurv}. The best 
parametrization that we have found is a linear function times the logarithm
of $l$. A more detailed analysis shows that this new parametrization reduces
the square root of the sum of the square of the relative differences 
((fit-data)/data)
by one order of magnitude. This suggests that one should try
to derive rigorous bounds where the errors are proportional
to some inverse power of $(l_{max}!)$. 

We have thus studied the logarithm of
the relative differences (due to the change in $l_{max}$) divided by
the logarithm of $l_{max}$ for various temperatures with $D=3$. The results
are shown in Fig. 7. We then used linear fits for the part 
falling linearly. In other words, we assumed the approximate law 
\begin{equation}
|{{\chi^{l+1}-\chi^l}\over{\chi^l}}|\simeq l^{(-|s|l+q)} \ .
\end{equation}
The results can be summarized as follows. The slope
is almost independent of $\beta $ and takes the approximate value -0.41 with
changes of order 0.01. The intercept grows linearly with 
$-Log_{10}(\beta _c -\beta )$,
as shown in Fig. \ref{intercept}. 
A linear fit of this data gives an 
intercept of the form $1.7-0.83 \times Log_{10}(\beta _c -\beta )$.
If we neglect the slow logarithmic variations and approximate it by a constant
central value in the falling part of Fig. \ref{ldep3}, 
we obtain the approximate law
\begin{equation}
|{{\delta \chi }\over{\chi}}|\sim 3.2 \times 10^2
\times (\beta _c - \beta)^{-1.2}\times (4.1)^{-l_{max}} \ .
\end{equation}
If we require these errors to be smaller than the numerical errors, we find
that $l_{max}=40$ is a safe choice for all the values of $\beta_c -\beta$
accessible with double precision.
Slightly larger values are obtained for $D=4$ and 5, which confirms that 
the last column of Eq. (\ref{eq:crit}) represents 
approximately the values of $l_{max}$ above which
no significant changes are observed. In conclusion, for calculations 
using double precision, the choice $l_{max}=50$ is convenient and safe
for the three values of $D$ considered above.

Having an acceptable control on the susceptibility
guarantees that we have an acceptable control
on the higher moments, $<M_n^{2q}>_n /2^{qn}$ for $q>1$, 
since to leading order in the 
volume, these quantities are dominated by the disconnected parts.
The precision which can
be achieved on the connected parts (which enter in the definition
of the renormalized coupling constants) is a more delicate question, which 
is discussed in section \ref{sec:lam}.

\section{Effects of Finite Dimensional Truncations on the HT Coefficients}
\label{sec:ht}

In a previous publication\cite{osc}, we used the 
truncated algorithm to calculate 800 coefficients
of the high-temperature expansion of the magnetic
susceptibility. We claimed that this  
truncation did not affect the numerical values
obtained. In this section, we provide a more systematic 
justification of this procedure.

We examine the $l_{max}$-dependence of the high-temperature
coefficients of the susceptibility, for dimensions 3, 4, and 5.  As in
section ~\ref{sec:ldep}, we replace $l_{max}$ with $l$ for notational 
purposes. We denote the
high-temperature coefficients as $b_{m}$ for the $m$th coefficient.

For $l$ and $m$ large enough, we find good linear fits in $l$ for the 
quantity 
\[
\frac {Ln( {( b_{m}-b_{m}^{(l)})}/{ b_{m} } ) }{Ln(l)} \ ,
\]
where $b_{m}^{(l)}$ is the truncated version
of the exact $b_{m}$. Fig. \ref{bldepd3} 
show these lines in $D=3$, for $m$ = 200, 300,
and 400. Very similar numerical values 
are obtained for $D=4$ and $D=5$. For this reason,
it was impossible to display the values for the three chosen dimensions in
a single graph. The graphs in $D=4$ and $D=5$ are similar looking and show
a linear behavior as good as in $D=3$. 
We can thus express the truncated coefficients as
\begin{equation}
b_{m}^{(l)}=b_{m} (1-l^{-|s|l + q}) ,
\end{equation}
where $s$ and $q$ are, respectively, the slope and intercept of the
corresponding fitted line.  
For the three chosen dimensions, the lines seem to ``focus'' in one 
point close to the $l=0$ axis. 
The intercepts are approximately independent 
of $m$ and take the approximate values 3.4 ($D=3$), 2.3 ($D=4$),
and 1.7 ($D=5$).
For the slopes, we find straight line fits if we plot 
\[
 \frac{Ln(-s(m))}{Ln(m)}
\]
versus $Ln(m)$ for $D=3$, and versus $m$ for $D=$ 4 and 5.  In 
Figs. \ref{slope3} and \ref{slope4}
we have used every tenth coefficient in the range from $m=300...400$.
From these fits, we find for $D=3$:
\begin{equation}
s = - m^{.013 Ln(m) - .22}.
\end{equation}
For $D=4$ we found
\begin{equation}
s = -m^{.000024 m-.19} ,
\end{equation}
and for $D=5$:
\begin{equation}
s=-m^{.000028 m-.21} .
\end{equation}

In $D=3$, for example, these results show us that we need $l$
larger than 34 for the error on $b_{1000}$ to be less than $10^{-16}$,
while for $D=4$, $l>38$.  Therefore, the value
of $l=50$ we have previously used in calculating the first 800
coefficients is more than adequate.

\section{Calculation of Subtracted Quantities}
\label{sec:lam}

In this section, we discuss the numerical aspects of the 
calculation of subtracted quantities.
We specialize the discussion to the calculation of the ``dimensionless 
renormalized coupling constant''\cite{parisi} corresponding
to the 4-point function. 

From Eq.(\ref{eq:moments}), it is clear that the calculation of $R_n (k)$
allows us to determine the renormalized coupling constants.
The first step in the calculation of these quantities is
to extract the connected parts. 
In other words, we first subtract the disconnected parts from the $2k$-point
function.
From a numerical point of view,
this is not a trivial operation, because the 
subtracted quantities (connected parts) 
scale differently
with the volume than the parts of which they are made.
For instance, for $\beta < \beta _c$ and $n$ sufficiently large,
\begin{equation}
<M_n^4>_n-3(<M_n>_n)^2 \propto 2^n \ ,
\label{eq:conn} \end{equation}
while the individual components scale like $2^{2n}$. The situation
is worse if we consider the 6-point functions, where the connected
part has the same scaling as Eq. (\ref{eq:conn}) but
the individual components scale like $2^{3n}$.
In other words, the beginning significant numbers of the individual
terms do not matter for the subtracted quantities. 
Assuming 16 significant digits, when $2^{2n}$ reaches $10^{16}$, we still
get the subtracted parts with 8 significant digits. When $2^n$ reaches 
$10^{16}$, there are no significant digits left for the subtracted part.

As a consequence, it is not always possible to stabilize the value
of the connected part during as many iterations as we would like, 
given the study of section \ref{sec:vol}. This is 
an interesting situation. As long as we increase the number of iterations,
we get a value of the unsubtracted quantity which becomes closer to
its infinite volume limit. If we represent the 
significant digits of a double precision number 
as a sequence of 16 digits written in the conventional way,
we can visualize this procedure as the successive
obtention of the digits on the right side of the number.
Unfortunately, at the same time, the part which gets subtracted increases 
in magnitude. Consequently, more an more digits on the left side of the
number are wasted for the evaluation of the subtracted quantities.
The situation gets worse if we consider 
the 6 or higher point functions.

The subtracted quantities are diverging near criticality. However,
it is possible to define\cite{parisi}
dimensionless renormalized coupling constants which have a finite limit.
In the case of the four point function, the 
dimensionless renormalized coupling constant
$\lambda _4$ is obtained 
by multiplying the zero-momentum 
connected Green's function $G_4^c$
by the $D+4$ power of the renormalized mass $m_R$, namely
\begin{equation}
\lambda _4 =- G_4^c m_R^{D+4} \ .
\label{eq:lam} \end{equation}
The mnemonic for $D+4$ is 8 (amputation of the 4 legs at zero-momentum) +
$D-4$ (the canonical dimension of the $\phi^4$ bare coupling constant).
We are using the notation
\begin{equation}
G_4^c=lim _{n\rightarrow \infty} {{<M_n^4>_n-3(<M_n>_n)^3} \over {2^n}} 
\end{equation}

In order to compare with field theory results, one
should consider Landau-Ginzburg measures where the 
cut-off dependence has been restored explicitly. 
For instance, in $D=3$, the definition of
the functions entering in the Callan-Symanzik equations 
(the beta function etc.) given in Ref.\cite{parisi} requires
that we keep the {\it dimensionful} constant fixed while the cut-off goes 
to infinity. In other words, we need to change 
the {\it dimensionless} constant
entering in $R_0(k)$ while taking the infinite cut-off limit. This delicate
procedure is beyond the scope of this paper, where we emphasize the basic
numerical aspect of a single calculation. As explained in the introduction,
we continue using a fixed Ising measure
and a single adjustable parameter ($\beta $). 

The quantity $\lambda_4$ has
a finite (and supposedly non-zero when $D<4$) limit
when $\Lambda \rightarrow \infty$ or equivalently when 
$\beta \rightarrow \beta_c$. We can thus bypass the explicit introduction
of the cut-off. Taking into account that there is no
wave function renormalization, or in other words that
the critical exponent $\eta$ is zero, we define $\lambda_4$
as the limit where $n$ goes to $\infty$ of 
\begin{equation}
\lambda_{4,n}= {{{<M_n^4>_n-3(<M_n^2>_n)^2}}
\over{{2^n}({{<M_n^2>_n} \over {2^n}})^{{D\over2}+2}}} \ .
\end{equation}

Equivalently, with the convention of Eq. (\ref{eq:rn}), which does not
involve inverse factorials, and for the value 
$s={2 {c^{-{1\over 2}}}}$ always used here, we obtain
\begin{equation}
\lambda_{4,n}= 12{{a_{n,1}^2 -2 a_{n,2}}\over{(-2a_{n,1})^{{D\over2}+2}}}\ .
\end{equation}

In practice, we pick a given relative precision $\Delta$ and we require
that $n$ is large enough to stabilize the susceptibility {\it and}
$\lambda_4$ with a relative precision $\delta$. 
The reason for requiring both conditions is that $\lambda_4$ may temporarily
stabilize when the flow passes near the fixed point (and so we are still
far away from the infinite volume limit), but this is signaled by
the fact that the susceptibility is still growing.
In summary, we require
\begin{equation}
|{a_{n+1,1}\over a_{n,1}}-{c\over 2}|<\Delta
\end{equation}
and 
\begin{equation}
|{\lambda_{4,n+1}\over \lambda_{4,n}}-1|<\Delta \ .
\end{equation}
When these two conditions are satisfied, we check that our result for
$\lambda_4$ is compatible with the expected precision, or in other words,
that we have enough significant digits left in $a_{n,1}^2 -2 a_{n,2}$ to 
calculate $\lambda_4$ with a relative precision $\Delta$. We thus require
the additional condition
\begin{equation}
|{{a_{n,1}^2 -2 a_{n,2}}\over{2a_{n,2}}}|>{\delta \over \Delta} \ ,
\end{equation}
where $\delta$ is a typical round-off error ($10^{-16}$ in double
precision).
If the additional condition is not satisfied, 
we lower $\delta$ and repeat the calculation.
We have applied this algorithm in $D=3$, 4, and 5 and for $-Log_{10}(\beta_c-
\beta)=2, 3,\dots , 14$.

For $D=3$, we were able
to do all the calculations with $\Delta=10^{-6}$. We found that 
$\lambda_4$ reaches a limit $\lambda_4^{\ast}=1.92786$
when $\beta \rightarrow \beta_c$. 
In other words, hyperscaling holds very well.
Fig. \ref{ren3}
shows that to a good approximation 
\begin{equation}
\lambda_4 -\lambda_4^{\ast}\simeq 1.68\times (\beta_c-\beta)^{-0.43} \ .
\end{equation}

In $D=4$, we had to reduce to  $\Delta=10^{-4}$. We found that
$\lambda_4$ tends to zero
when $\beta \rightarrow \beta_c$.
As shown in Fig. ~\ref{ren4}, we have the approximate law
\begin{equation}
\lambda_4 \simeq {1\over{-1.96-0.746\times Ln(\beta_c-\beta)}} \ ,
\end{equation}
which is consistent with perturbative calculations\cite{brezin}.

In $D=5$, we had to reduce further to  $\Delta=10^{-2}$. We found that
$\lambda_4$ tends to zero
according to the approximate law
\begin{equation}
\lambda_4 \simeq 1.02\times (\beta_c-\beta)^{0.507} \ ,
\end{equation}
as shown in Fig. \ref{ren5}.
If we replace $(\beta_c-\beta)$ by $\Lambda^{-2}$, we see that our 
result is consistent with $\lambda_4 \propto\Lambda^{-1}$.  

\section{Conclusions and Perspectives}

We have shown that the use of truncations in the Fourier transform 
of the recursion relation of Dyson's hierarchical model leads
to systematic errors which can be suppressed more than exponentially
when the dimension of the truncated space increases.
We have justified the use of the truncation 
for calculations\cite{osc} of the high-temperature expansion. 
We have shown that the finite volume effects 
can be reduced with an exponential precision.
We have 
found the temperature dependence of the numerical errors and
explained the empirical results with a simple renormalization group
argument. 

The numerical errors appear as a practical aspect of
the hierarchy problem. If it seems hard to 
believe that nature would fine-tune its fundamental parameters
to produce scalar particles with masses very small compared to the 
Planck scale, there are also practical difficulties related 
to this fine-tuning.
In the present context, it is quite simple to find $\beta _c $. However,
this is not the end of the story: the physical quantities become 
numerically unstable when we reach $\beta _c$.  This difficulty
is not unsurmountable if we want to reach a cut-off of the order of the Planck
scale which is {\it only } 17 orders of magnitude larger
than the weak scale. We can use programming methods with enough significant
digits. To take the analogy with differential equations,
the problem of sensitive dependence on initial conditions can be dealt
with provided that we do not evolve the system during a too-long amount
of time. 

The methods presented here can be applied to  
field theoretical calculations. The simplest one is
the calculation of the renormalized mass.
This quantity is crucial because 
it enters into the definition of the functions appearing 
in the Callan-Symanzik equations\cite{parisi}.
For the model
considered here, a possible definition  
of the renormalized mass is
\begin{equation}
m^2_R (\mu )=lim_{L\rightarrow \infty} 
{\Lambda ^2 \over {\chi_{\infty}(\beta _c  +\lambda ^{-L} \mu)}} \ ,
\end{equation}
where $\lambda $ is the largest eigenvalue of 
the linearized renormalization group
transformation which needs to be calculated precisely. 
$\Lambda $ is a UV cut-off taking the value $2^{{L\over D}} \Lambda _R$ 
where $\Lambda _R$ is a scale of reference below which we are considering 
an effective theory.
Finally, $\mu $ is a 
parameter which allows us to
change the value of the renormalized mass.
As explained in section 
\ref{sec:lam}, the method can also be applied to the calculation of 
renormalized quantities in Landau-Ginzburg models. These calculations
will be used to check the validity of the perturbative evaluations of
the functions entering the Callan-Symanzik equations.
In the case $D=4$, the high-temperature expansion\cite{ht4} indicates that
the perturbative result\cite{brezin} is very accurate.

More generally, the calculational method presented here can be used
to check any kind of approximate calculation which applies to 
the hierarchical model.

\acknowledgments

This research was supported in part by the Department of Energy
under Contract No. FG02-91ER40664.

\newpage
\begin{figure}
\vskip30pt
\centerline{\psfig{figure=invbetc.ps,height=5in,angle=270}}
\vskip50pt
\caption{ $1\over{\beta _c}$ versus the dimension $D$. }
\label{invbetc}
\end{figure}
\newpage
\begin{figure}
\vskip20pt 
\centerline{\psfig{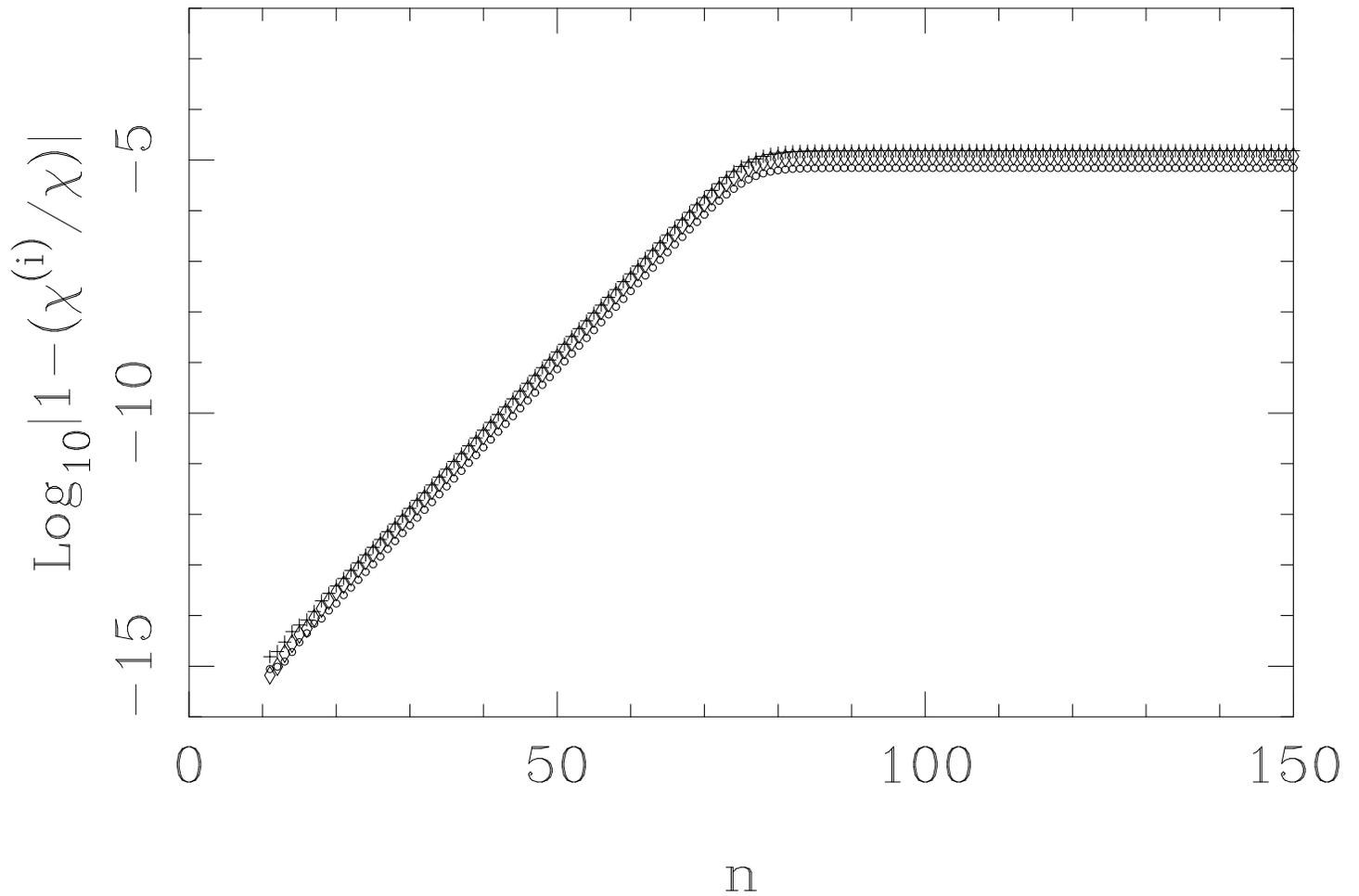}}
\vskip50pt
\caption{Relatives differences between the susceptibility calculated with the 
main method $(\chi)$ and three alternative methods 
($\chi^{(i)}$) with $i=1$ (crosses), 2 (diamonds) and 3 (circles) 
as in the text. The calculations were done in $D=3$ and with 
$\beta=\beta_c-10^{-11}$.}
\label{3err}
\end{figure}
\newpage
\begin{figure}
\vskip20pt
\centerline{\psfig{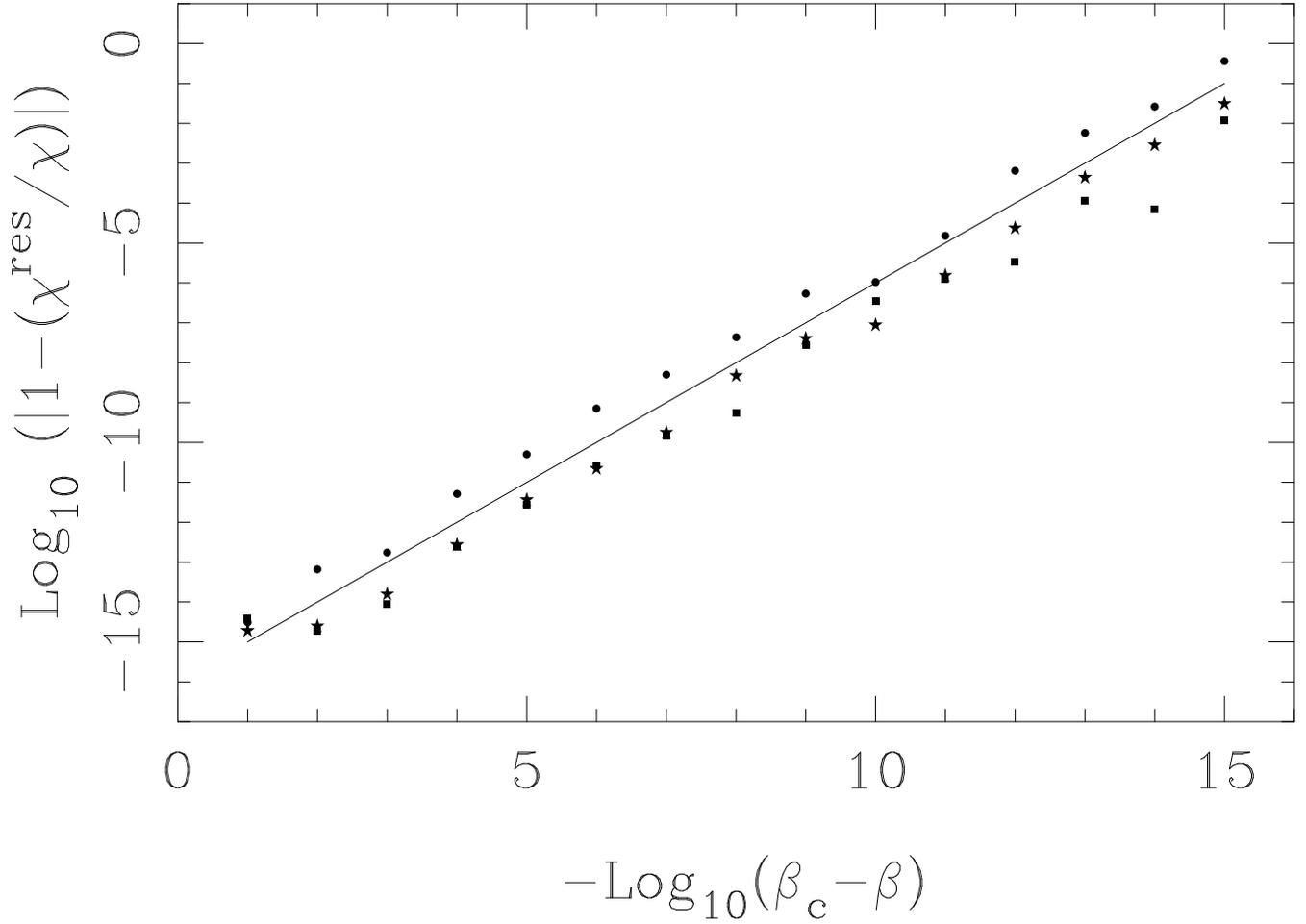}}
\vskip50pt
\caption{Relative difference between the susceptibility calculated with the 
main method $\chi$ and with a rescaling $\chi^{res}$ as explained in the text
versus $\beta_c-\beta$, in $D=3$ (circles), $D=4$ (stars) and $D=5$ (squares).}
\label{dsus345}
\end{figure}
\newpage
\begin{figure}
\vskip20pt
\centerline{\psfig{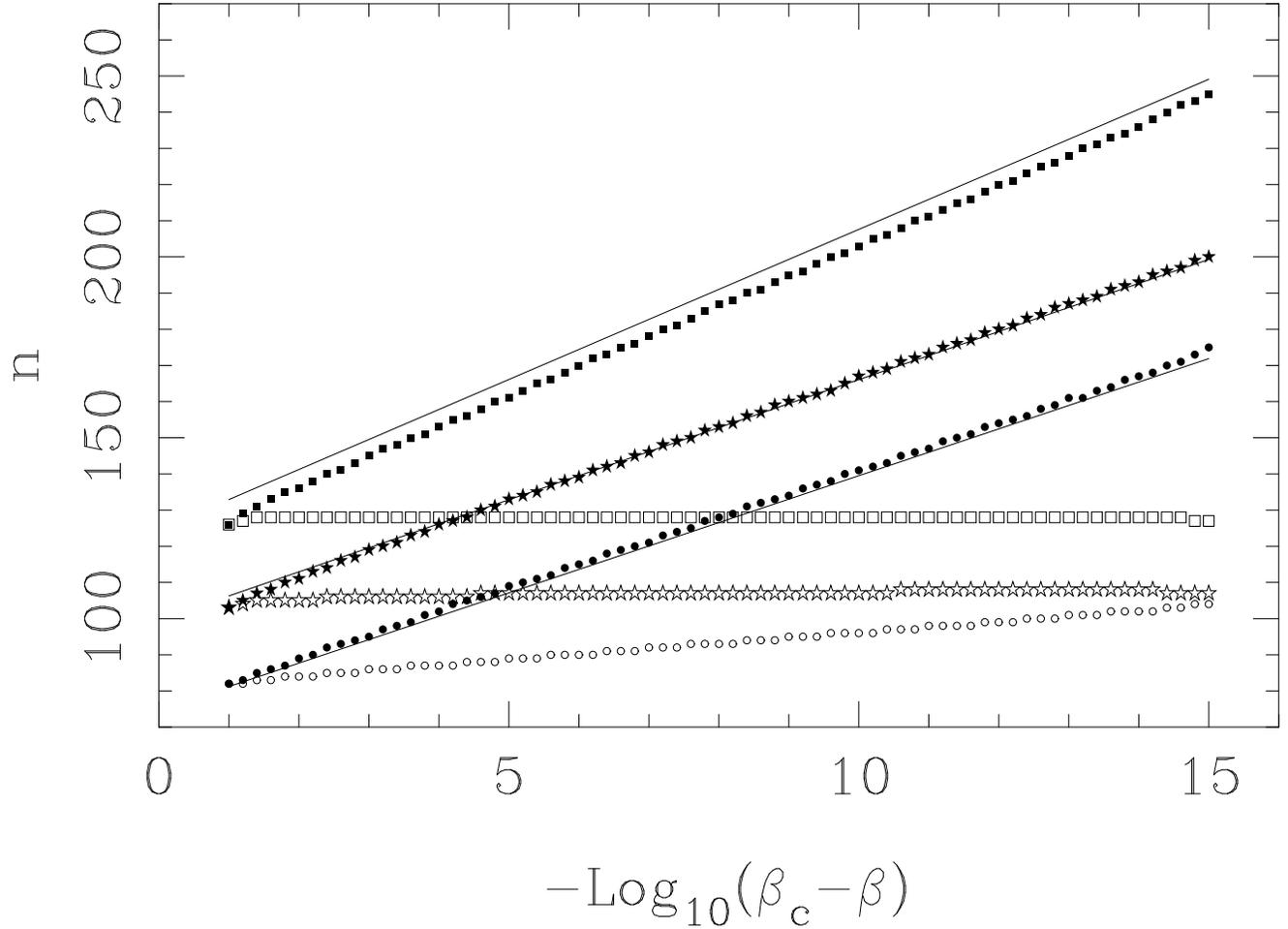}}
\vskip50pt
\caption{$n(\beta,\Delta)$ defined in Eq.(\ref{eq:nbeta}), 
for $D=3$ (circles), $D=4$ (stars) and $D=5$ (squares). Filled symbols 
correspond to a fixed value $\Delta=10^{-15}$, empty symbols correspond
to a variable value $\Delta=10^{-16}/(\beta_c -\beta)$.}
\label{vol}
\end{figure}
\newpage
\begin{figure}
\vskip20pt
\centerline{\psfig{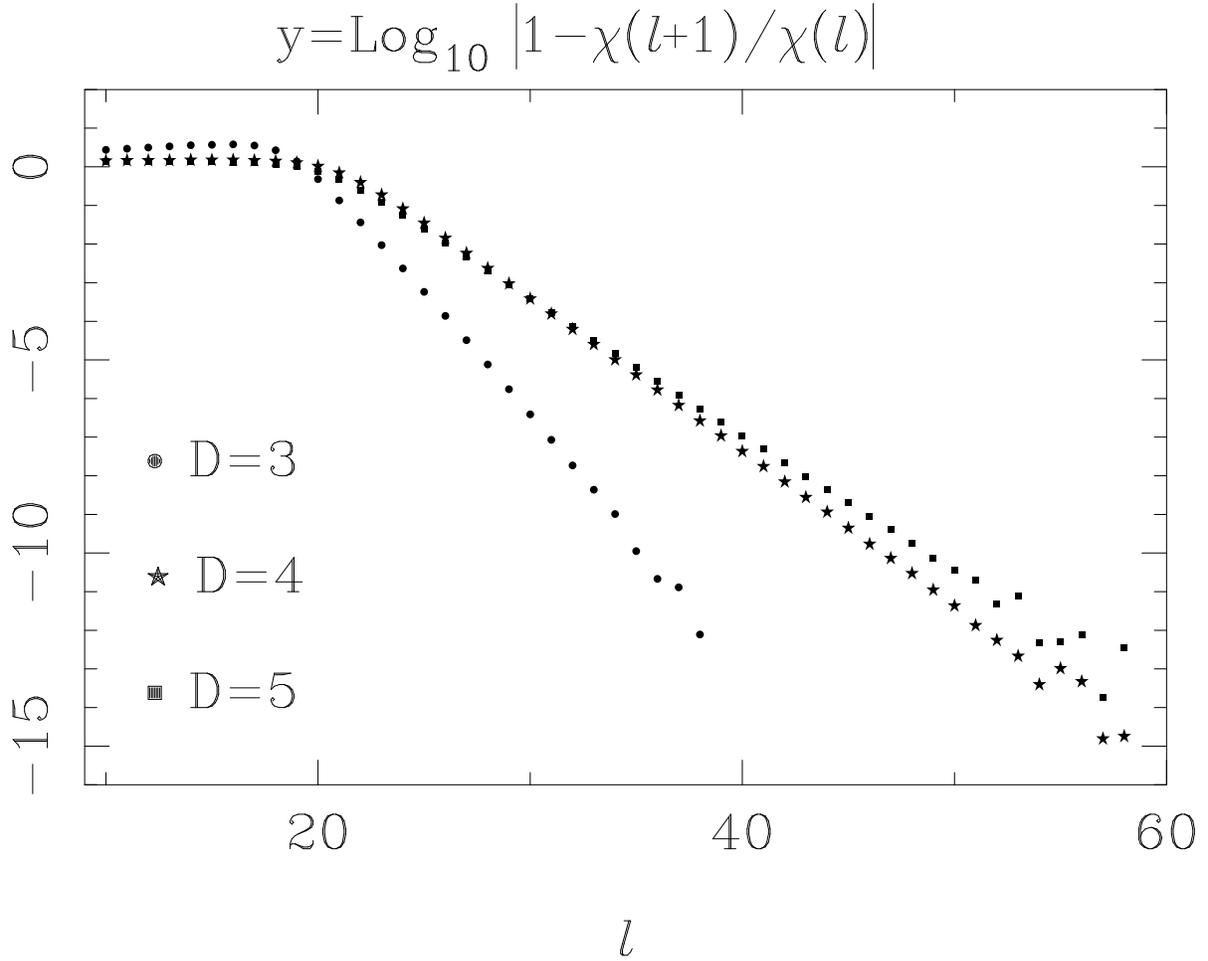}}
\vskip50pt
\caption{Relative difference between the susceptibility calculated with
$l_{max}=l+1$ and $l_{max}=l$ 
in $D=3$ (circles), $D=4$ (stars) and $D=5$ (squares). 
$\beta=\beta_c -10^{-8}$
in the three cases.}
\label{ldep}
\end{figure}
\newpage
\begin{figure}
\vskip20pt
\centerline{\psfig{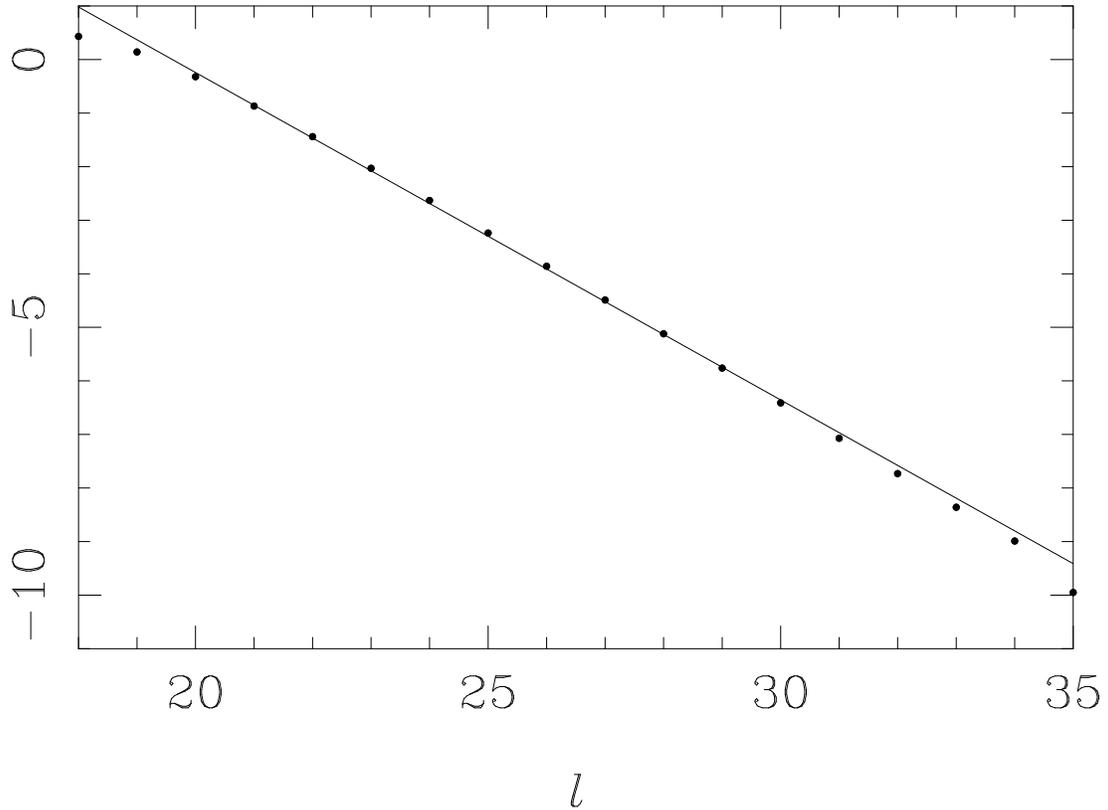}}
\vskip50pt
\caption{Relative difference between the susceptibility calculated with
$l_{max}=l+1$ and $l_{max}=l$ 
in $D=3$ (circles) and with $\beta=\beta_c -10^{-8}$, compared with  a linear
fit of these points.}
\label{ldepcurv}
\end{figure}
\newpage
\begin{figure}
\vskip20pt
\centerline{\psfig{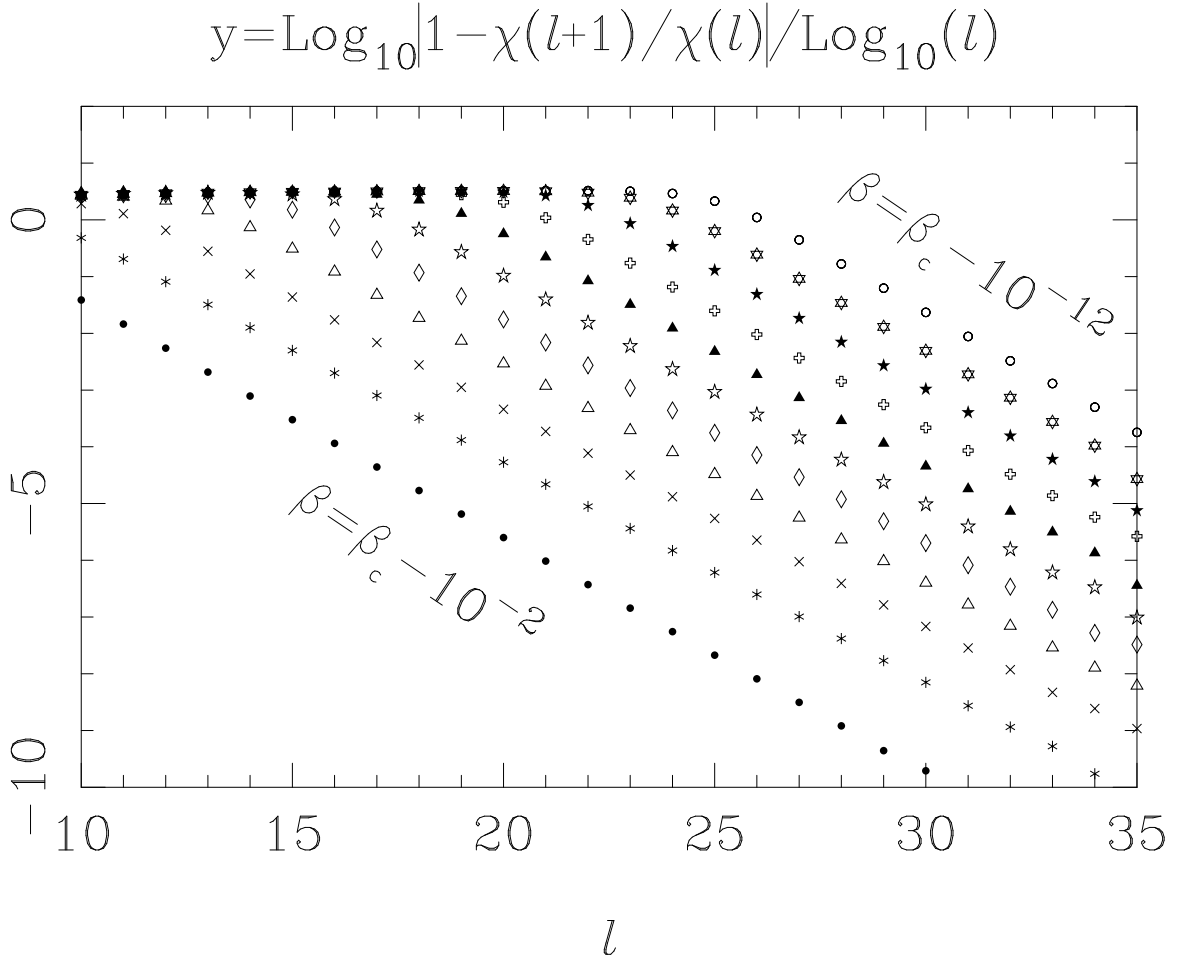}}
\vskip50pt
\caption{
Relative difference between the susceptibility calculated with
$l_{max}=l+1$ and $l_{max}=l$ 
in $D=3$ with $\beta=\beta_c -10^{-2}$ (filled circles), 
$\beta=\beta_c -10^{-3}$ (asterisques), 
$\beta=\beta_c -10^{-4}$ (crosses),......up to $\beta=\beta_c -10^{-12}$ 
(empty circles).
}
\label{ldep3}
\end{figure}
\newpage
\begin{figure}
\vskip20pt
\centerline{\psfig{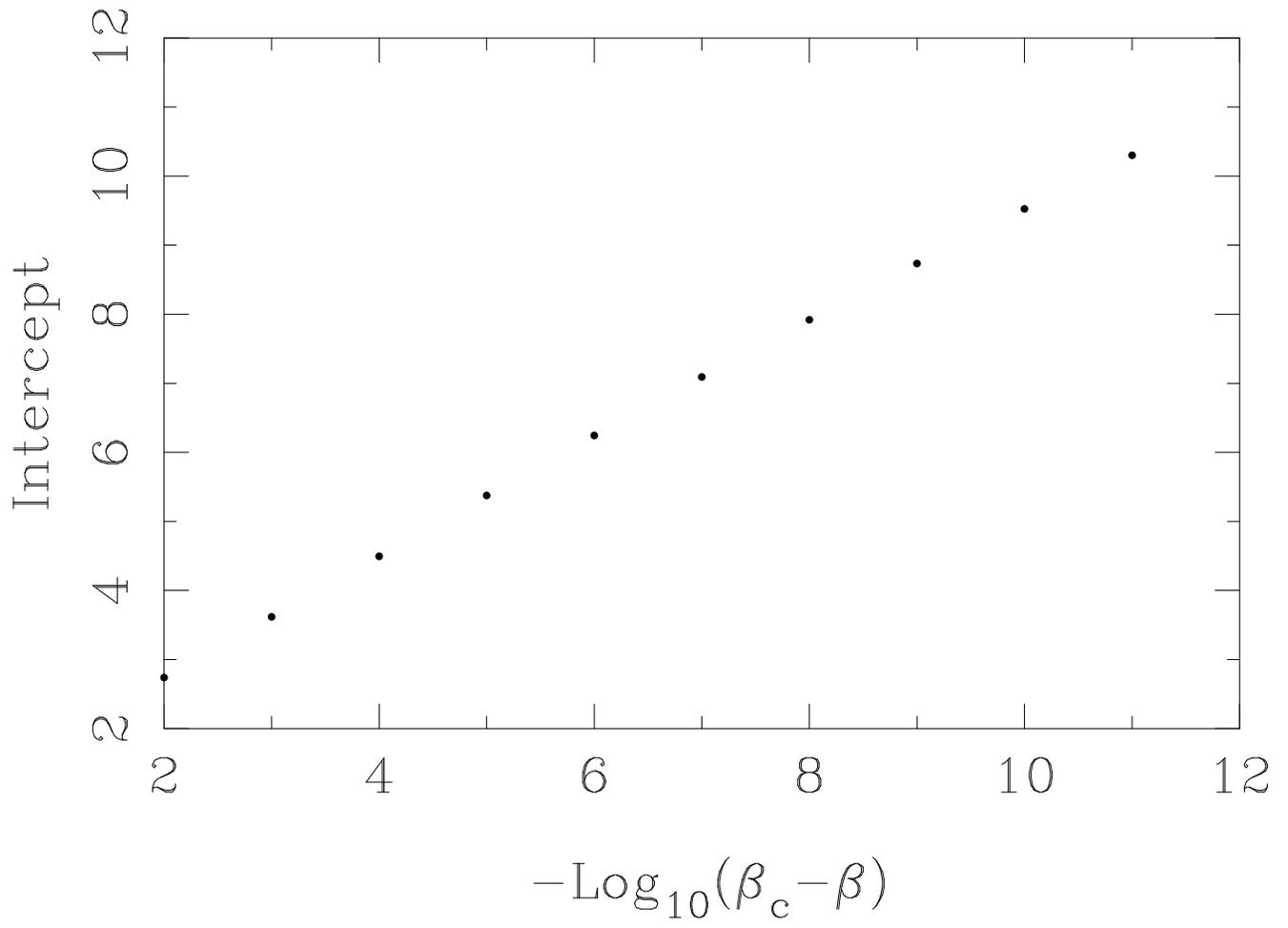}}
\vskip50pt
\caption{Intercept of the linear fits corresponding to the linear part
of Fig. ~\ref{ldep3}.}
\label{intercept}
\end{figure}
\newpage
\begin{figure}
\vskip20pt
\centerline{\psfig{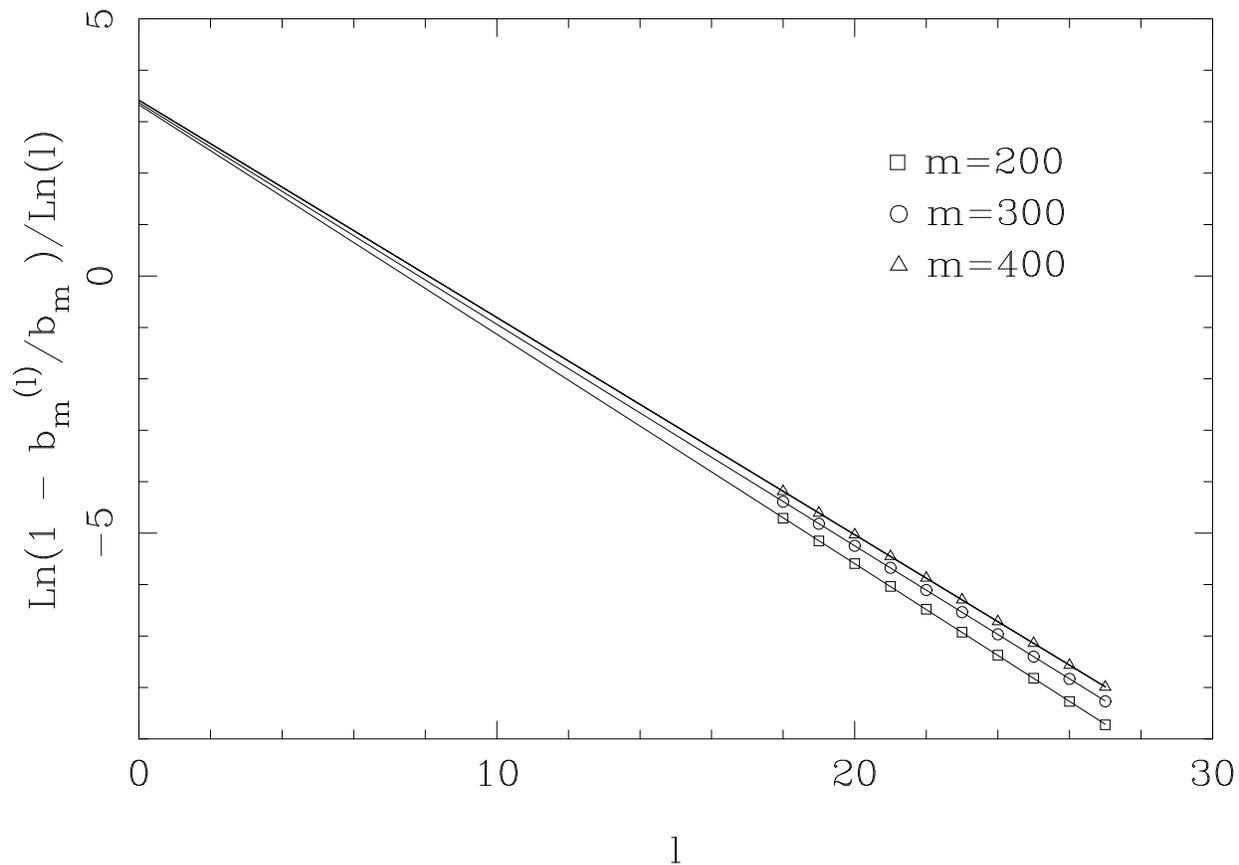}}
\vskip50pt
\caption{Relative difference between the coefficients of the 
susceptibility $b_m$ calculated with
$l_{max}=l+1$ and $l_{max}=l$ 
in $D=3$ with m=200 (squares), m=300 (circles) and m=400 (triangles).
}
\label{bldepd3}
\end{figure}
\newpage
\begin{figure}
\vskip20pt
\centerline{\psfig{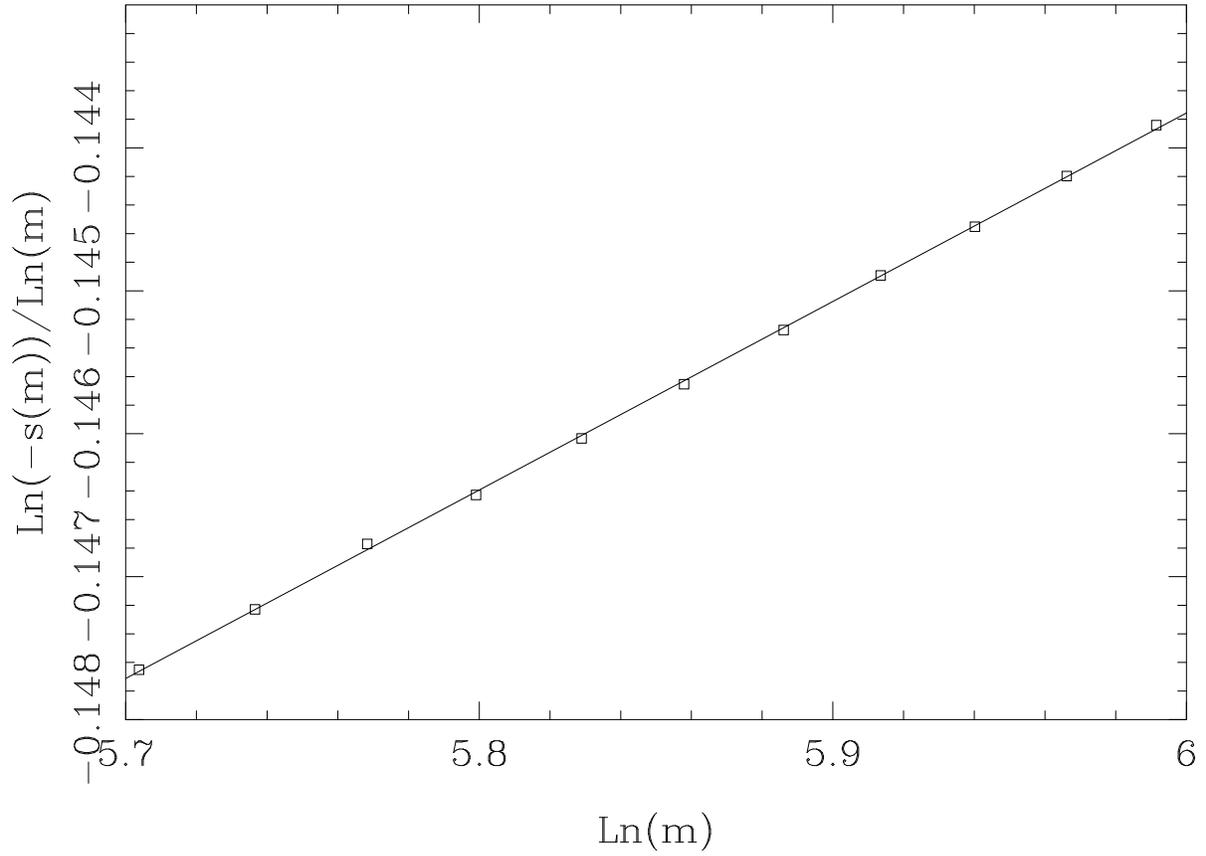}}
\vskip50pt
\caption{$Ln(-s(m))/Ln(m)$ versus $Ln(m)$ in $D=3$. }
\label{slope3}
\end{figure}
\newpage
\begin{figure}
\vskip20pt
\centerline{\psfig{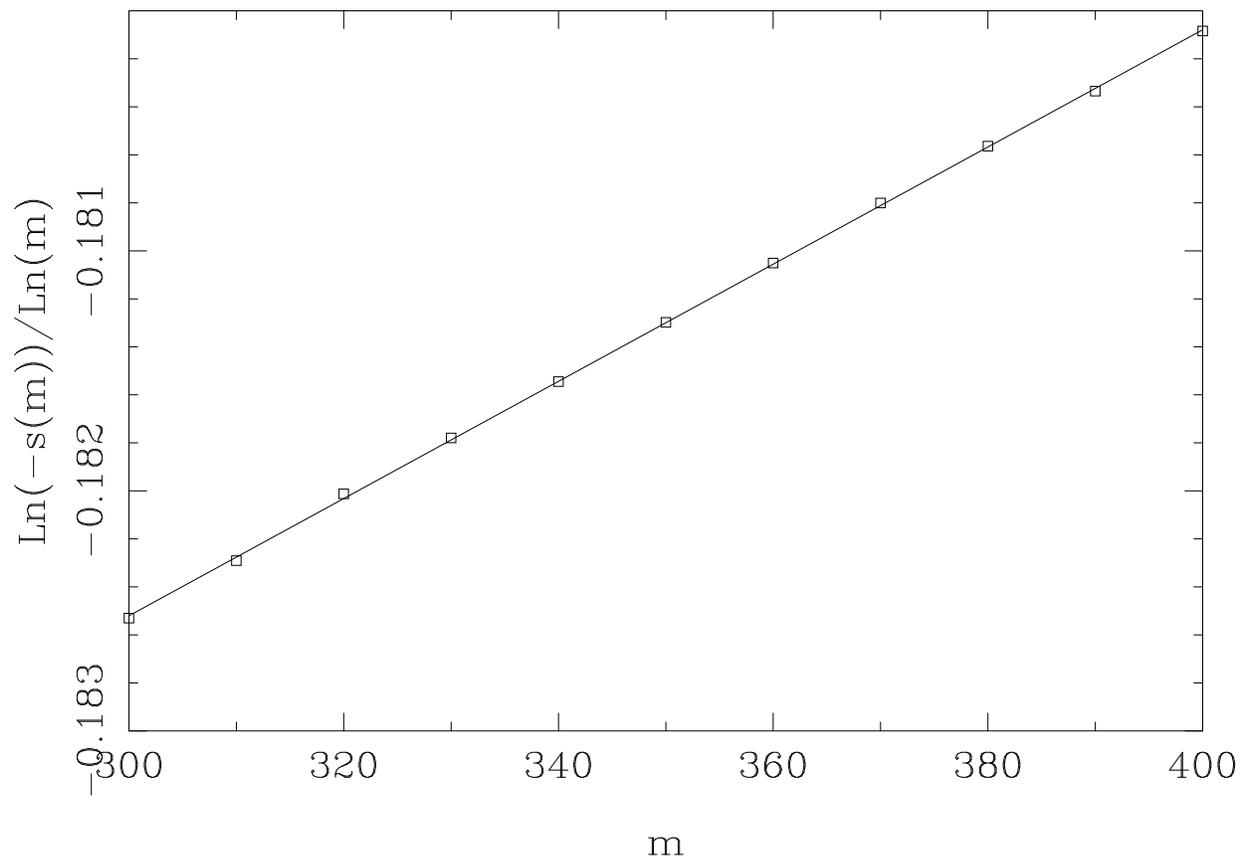}}
\vskip50pt
\caption{$Ln(-s(m))/Ln(m)$ versus $m$ in $D=4$.}
\label{slope4}
\end{figure}
\newpage
\begin{figure}
\vskip20pt
\centerline{\psfig{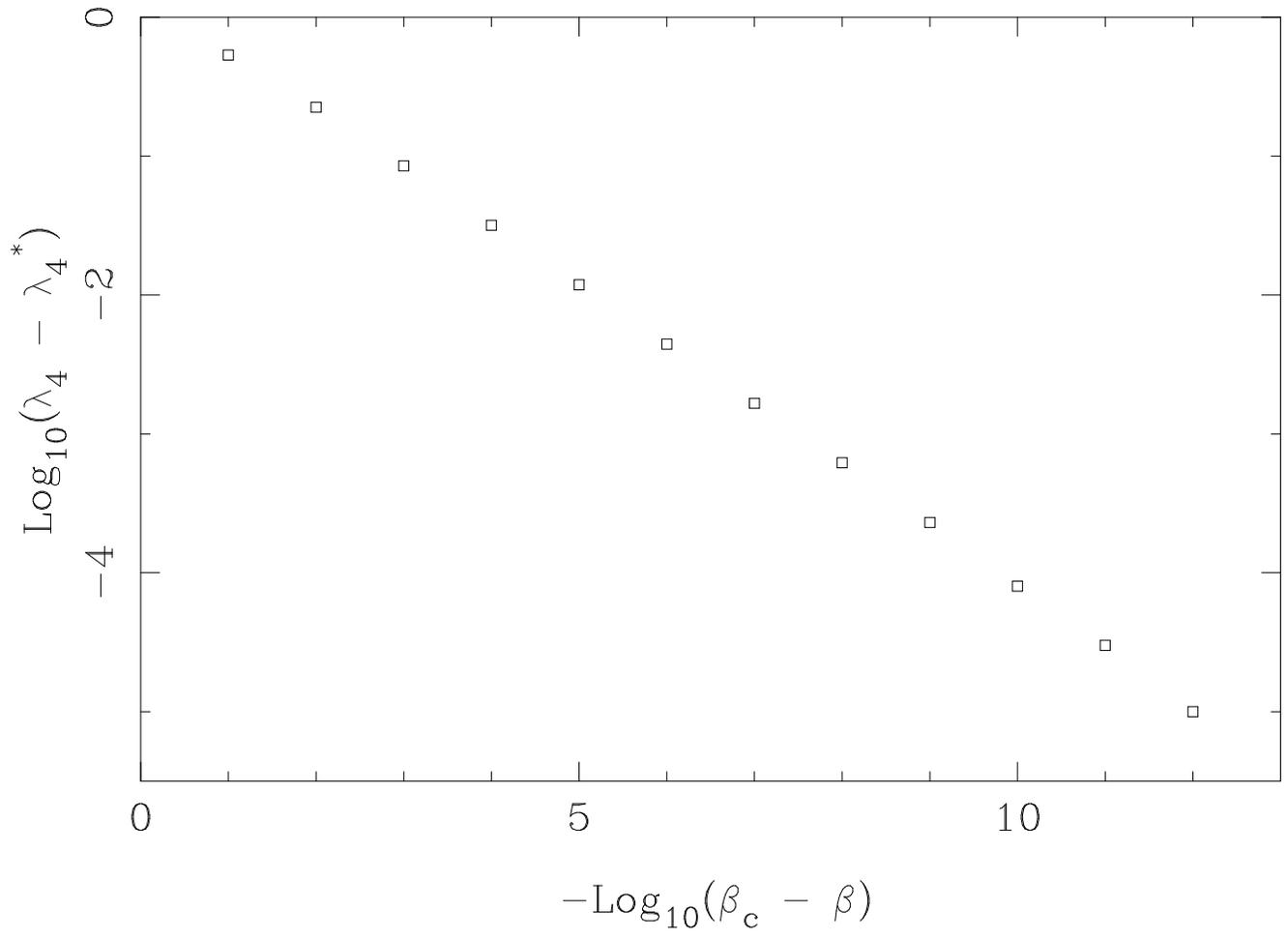}}
\vskip50pt
\caption{$Log_{10}(\lambda_4-\lambda_4 ^{\ast})$ 
versus $-Log_{10}(\beta_c -\beta)$ in 3 dimensions.}
\label{ren3}
\end{figure}
\newpage
\begin{figure}
\vskip20pt
\centerline{\psfig{figure=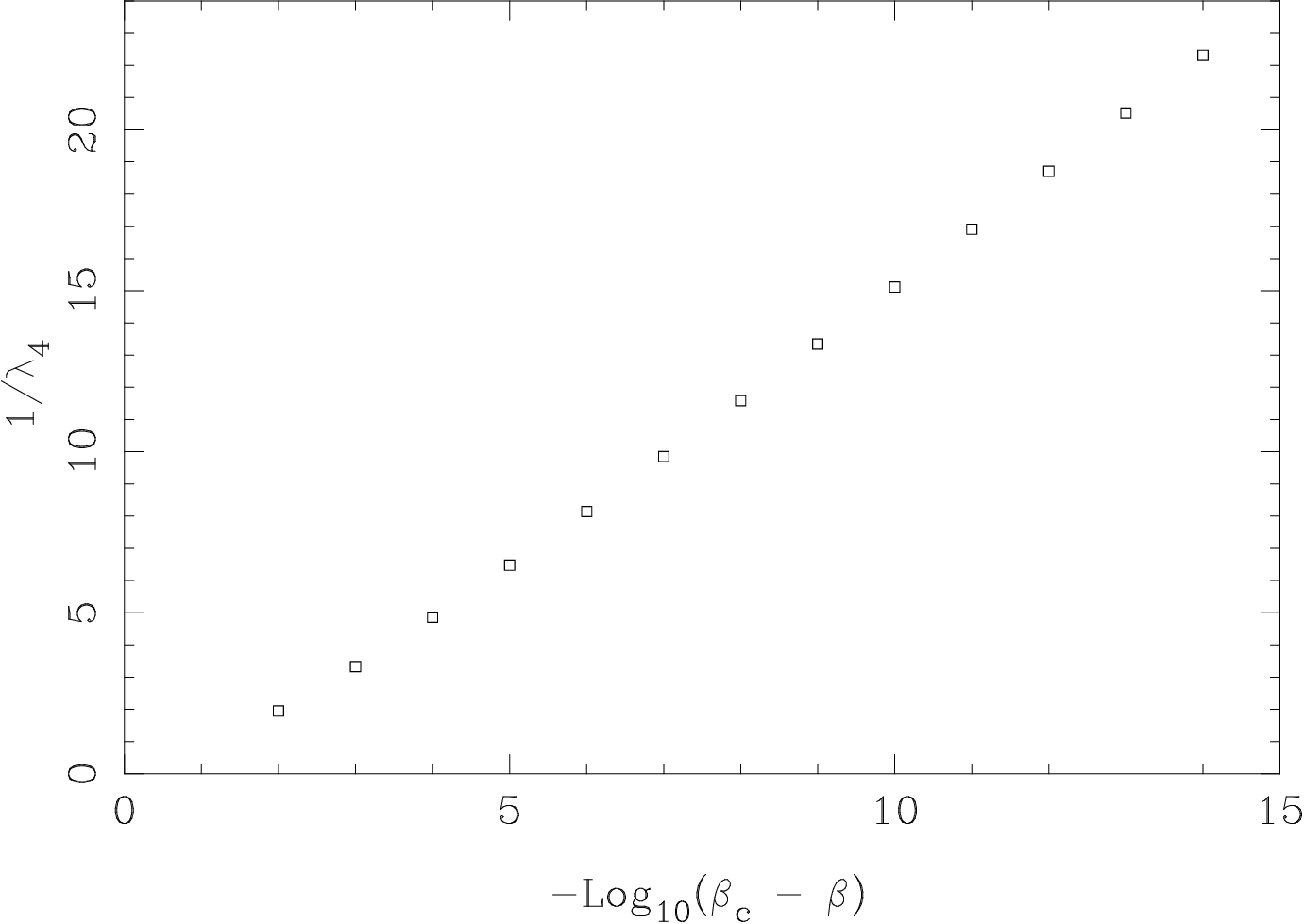,height=5in,angle=270}}
\vskip50pt
\caption{$1/\lambda_4$ versus $-Log_{10}(\beta_c -\beta)$ in 4 dimensions.}
\label{ren4}
\end{figure}
\newpage
\begin{figure}
\vskip20pt
\centerline{\psfig{figure=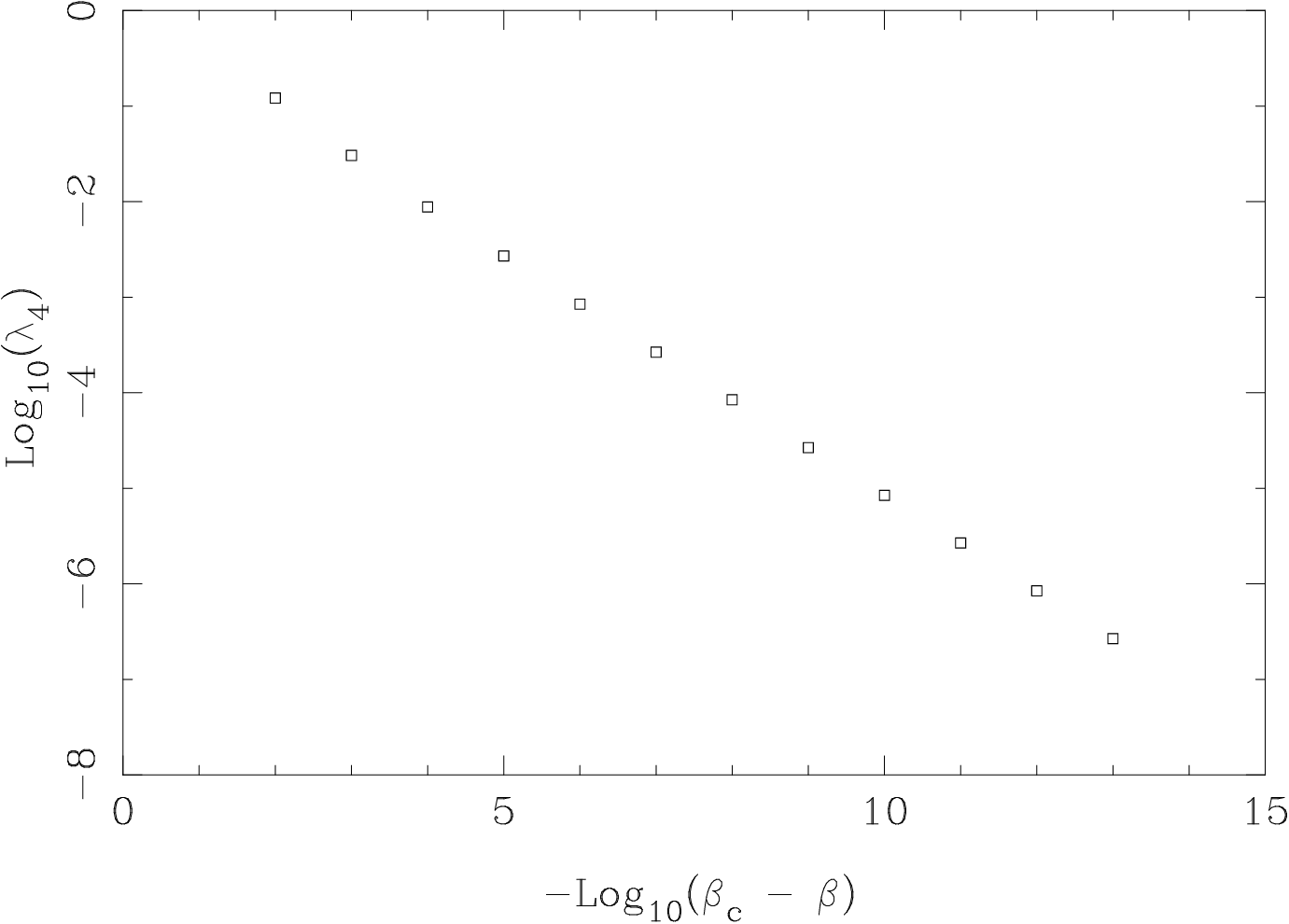,height=5in,angle=270}}
\vskip50pt
\caption{$Log_{10}(\lambda_4)$ 
versus $-Log_{10}(\beta_c -\beta)$ in 5 dimensions.}
\label{ren5}
\end{figure}
\end{document}